# Alignment Tolerant Broadband Compact Taper for Low-Loss Coupling to a Silicon-on-Insulator Photonic Wire Waveguide

**PURNIMA SETHI AND SHANKAR KUMAR SELVARAJA**

*[1]Centre for Nano Science and Engineering (CeNSE), Indian Institute of Science Bangalore, India.*
*\*shankarks@iisc.ac.in, purnimajosan@iisc.ac.in*

**Abstract:** We experimentally demonstrate a broadband, fabrication tolerant compact silicon waveguide taper (34.2 µm) in silicon-on-insulator. The taper works on multi-mode interference along the length of the taper. A single taper design has a broadband operation with coupling efficiency >70% over 700 nm that can be used in O, C and L-band. The compact taper is highly tolerant to fabrication variations; $\pm$100 nm change in the taper and end waveguide width varies the taper transmission by <5%. The footprint of the device i.e. taper along with the linear gratings is ~ 442 m$^2$; 11.5X smaller than the adiabatic taper. The taper with linear gratings provides comparable coupling efficiency as standardly used focusing gratings. We have also compared the translational and rotational alignment tolerance of the focusing grating with the linear grating.

## 1. Introduction

Photonics in silicon-on-insulator (SOI) platform is a tremendously promising technology due to its comprehensive applications owing to its compactness and CMOS-compatibility [1, 2]. Consequently, it has been the focus of a considerable research suited for enabling the integration of highly complex optical circuits for making compact devices. The strong light confinement in high index-contrast waveguide platform ensues dense optical integration with sub-micron dimensions and low bending losses which brings in new challenges in circuit design and routing [3-5].

The building block of an optical device/circuit is an optical waveguide which enables low-loss light propagation and is thereby, used to connect components and devices. Waveguides are generally designed with different cross-sections to realize various integrated photonic device such as, arrayed-waveguide gratings, spot-size converters, multimode interference couplers, grating couplers (GCs) as well as crossings [6-10]. The devices with different waveguide width should be connected through a low-loss interface. When footprint of these waveguide transitions. Since the taper length depends predominantly on the starting and ending waveguide width and the effective refractive index, the transition between a GC and a single-mode photonic waveguide in a SOI platform is substantial [11-15].

A grating footprint of 10 µm ×10 µm is typically chosen to mode-match the grating field with an optical fiber. The grating is then coupled to a waveguide through an adiabatic/non-adiabatic taper [16-25]. The function of the taper is to change the optical mode size and shape to achieve high coupling efficiency between the two waveguides of different cross-sections. In an adiabatic taper, the local first-order mode of the waveguide should propagate through the taper without coupling to higher-order modes and radiatiing modes. The adiabatic tapers in SOI wire waveguides are generally 300-500 µm long. Several designs of GC based adiabatic tapers have been proposed for SOI-based photonic devices, including linear [6, 16], exponential [9, 17], parabolic [18], and Gaussian [10]. However, the footprint of the spot-size converters based on linear GCs is limited by the length of the taper.

To reduce the footprint of the coupler, a compact focusing grating is commonly used which allows an 8X length reduction in the footprint (~18.5 µm × 28 µm) without performance penalty compared to a linear GC with an adiabatic taper [26]. However, focusing gratings require accurate fiber alignment, are bandwidth limited and suffer from reflection [27].

Thus, it would be exceedingly beneficial to use a linear GC with a short taper with low-insertion loss, low-reflection, broadband, alignment tolerant as well as robust to fabrication imperfections for a compact light-chip coupling scheme. Therefore, designing an improved waveguide taper for obtaining an efficient coupling between two different optical waveguides is essentially indispensable.

In the previous paper, we have shown an ultra-compact taper between a linear GC and single mode silicon waveguide using a quadratic sinusoidal function, merely 15 µm long with an insertion loss as low as 0.22 dB at 1550 nm and a bandwidth >150 nm [28]. However, the tapers were designed for shallow-etched waveguides. Wire waveguides allow low-loss sharp bends and thus, ultra-dense photonic circuits. In this paper, we propose a taper for wire waveguide with the combined advantages of broadband operation as well as compactness.

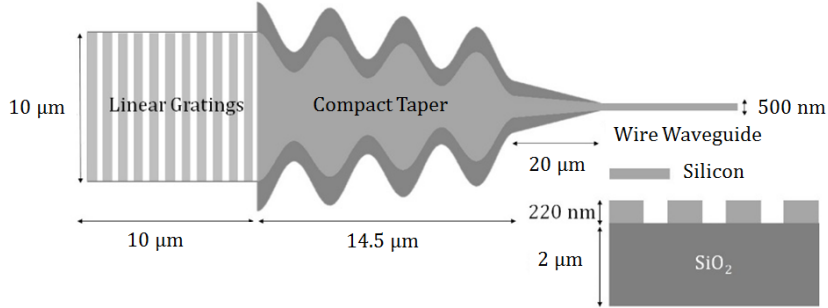

Fig. 1. Schematic of the compact taper structure for wire waveguide on SOI platform.

## 2. Compact Taper: Design and Simulation

The schematic of the taper structure along with GCs is shown in Fig. 1. The proposed taper works on self-imaging principle in multi-mode waveguides along the propagating length [28, 29]. The length and width of the taper are optimized to obtain interference progressively between the resonance modes along the taper resulting in maximum coupling to the fundamental waveguide mode. The interpolation formula used to define the proposed taper to connect a broad waveguide to a submicron waveguide section is [28, 29],

$$X = a\,[bz^2 + (1-b)z] + (1-a)[\sin^2\left(\frac{c\pi}{2}z\right)] \qquad (1)$$

where $a$ lies between 0 to 1, $b$ (used for fine tuning the optimal response) lies between -1 to 1, $c$ is any odd integer 3 ($c = 1$ creates the trivial case of half a sinusoidal oscillation). This formula meets the following boundary conditions: X ($z = 0$) = 0 and X ($z = 1$) = 1 where $z$ is the relative length of the taper. All four ($a, b, c, z$) design parameters allow one to design an appropriate taper profile for maximum transmission between the waveguides. The iterative feedback-based approach allows for lower design cycles as finer parameter spacing is required only near the optimum. The approach also greatly reduces the number of simulations, thereby reducing design time.

Table 1. Optimized parameters for the proposed taper

| Parameter | Value |
|---|---|
| Length of the Taper | 14.2 µm + 20 µm |
| 'a' variable | 0.4 |
| 'b' variable | 0.5 |
| 'c' variable | 7 |
| Width of Initial Waveguide | 10 µm |
| Width of Final Waveguide | 500 nm |
| Wavelength | 1.55 µm |
| Efficiency | 94.7% |

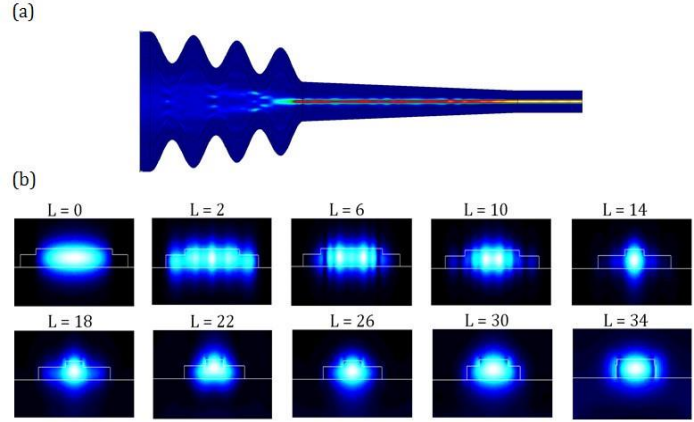

Fig. 2. Optical intensity profiles for the optimized compact taper at 1550 nm TE polarization (b) 2-D contour plot profile along the length (L) of the compact taper.

The simulations were carried out using Eigenmode Expansion (EME) propagation algorithm and overlap between modes is computed using Finite Difference Method (FDM). The compact taper was designed in an SOI wafer with 220 silicon on 2000 nm buffer oxide. Table 1 shows the optimized values of the relevant taper design parameters. A maximum coupling efficiency of 94.7% is achieved for a taper length of 34.2 µm shown by the optical intensity profile in Figure 2(a). Figure 2(b) illustrates the evolution of modes along the length (L = 34.2 µm) of the proposed taper.

Figure 3(a) shows the effect of end waveguide width variation from an optimized width of 500 nm on the transmission. The first four modes propagating in the waveguide are also shown in the inset. As is evident, > 75% transmission is achieved for a variation of ±200 nm. However, in practice one can expect a linewidth variation of < 10% which corresponds to a width change of ±25 nm. A variation in this range would result in transmission degradation by < 2% (0.08 dB), which shows the resilience of the proposed taper.

Figure 3(b) depicts the spectral response of the compact taper based on linear GCs. The inset shows the higher order modes for a wavelength span of 1000 nm. The proposed taper has a broadband operation with the 3dB bandwidth > 900 nm covering O, C, L-band and beyond. Furthermore, the effect of dimensional variation on the transmission performance was also calculated to take fabrication tolerances into account. Figure 3(c) shows the effect of the total taper width variation on the coupling efficiency obtained by varying the optimized *a* value. As is evident, the tapers are very resilient (> 80% efficiency for ± 500 nm shift in optimized taper width). During fabrication, a width variation of ±25 nm (25%) may occur, which results in a transmission degradation by merely < 1%. Figure 3(d) shows the effect of the total taper width variation on the coupling efficiency obtained by varying the optimized *b* value. The proposed structures have high manufacturing tolerances (> 80% efficiency with shift in optimized taper width of ± 200 nm).

Fig. 3. (a) Variation in taper's efficiency with waveguide width. Inset shows the variation for higher order modes, (b) Spectral response of the proposed compact taper in the C & L-band (1480 nm – 1640 nm) and beyond. Inset shows the broadband 1000 nm range for higher order

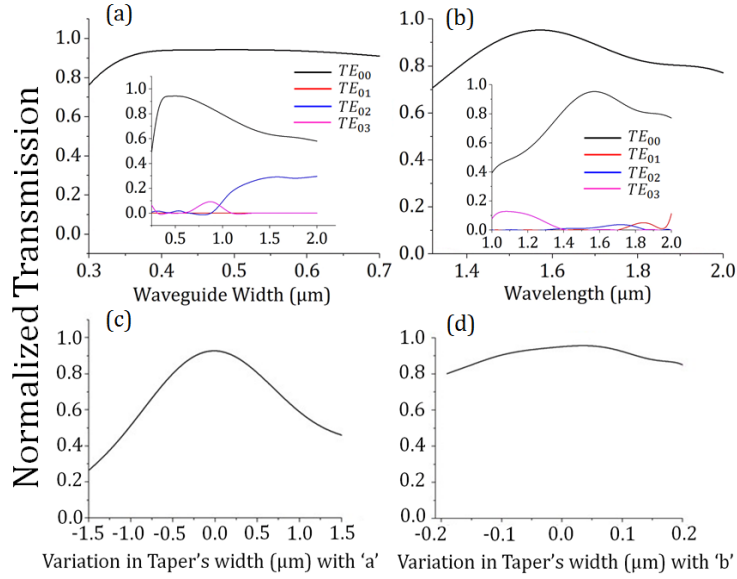

modes; Tolerance of the proposed taper i.e. effect of compact taper width variation with (c) 'a' and (d) 'b'.

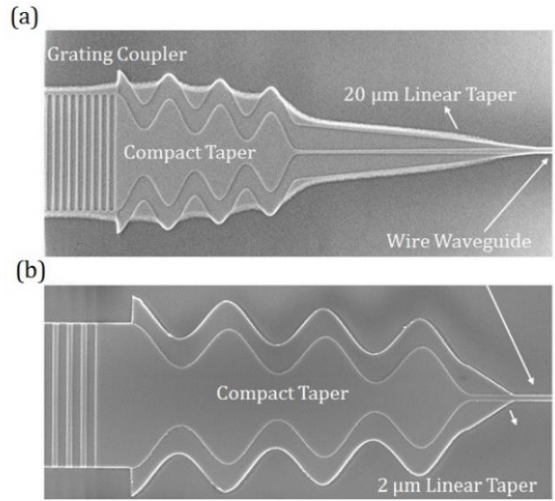

Fig. 4. Scanning Electron Microscope (SEM) of the proposed compact taper structure along an Adiabatic taper shoulder of length (a) 20 µm, (b) 2 µm.

## 3. Experimental Results and Discussion

To compare the proposed fiber-to-waveguide taper performance with the existing designs, three combinations of tapers were fabricated; (i) linear GCs with adiabatic taper, (ii) focused GCs and (iii) the proposed compact taper with linear GCs. The test structures were designed with an input GC coupling into a 500 nm wire waveguide and tapering-out to an identical output coupler configuration. All the GCs were designed for TE-polarized 1550 nm with grating period of 630 nm and 50% fill-factor [30].

The devices were fabricated using electron-beam lithography and Inductively Coupled Plasma-Reactive Ion Etching (ICP-RIE) process. Pattering was done on a standard SOI substrate with a 220 nm thick device layer on a 2 µm BOX layer. Figure 4 shows the scanning electron microscope (SEM) image of the proposed compact taper. Figure 4(a) shows the proposed structure along with a 20 µm long adiabatic taper. The taper shoulders along the wire waveguide aids in confining the mode. Figure 4(b) shows the proposed structure with a 2 µm long adiabatic taper. However, this configuration was less efficient, since the taper is short and hence, confinement is poor.

The fabricated devices were characterized using a tunable laser source (1510-1630 nm) and a photodetector. The polarization of the light from the laser source is controlled using polarization wheels before the input GC. The transmitted light is detected by an InGaAs photodetector. Figure 5 and Table 2 shows the summary of the characterization results. In order to see the tolerance to grating period variation, 5 set of devices were fabricated with period of 590, 610, 630, 650, and 670 nm.

Figure 5(a) compares the performance of the compact taper with the long tapers and focused GCs for different grating periods. The performance of the compact taper is marginally better than the long adiabatic taper. Although, focusing gratings are more efficient, their tolerance to fabrication imperfections is less. The efficiency of the focusing gratings, linear grating and compact taper degrades by 2.07/5.17, 1.98/3.58, 2.3/2.6 dB per coupler for a period shift of $\pm 40$ nm. The 3-dB bandwidth which is another important performance metric for a GC is ~ 5 nm higher for compact taper compared to a focusing GC for 630 nm period. The insertion loss per coupler is 6.2 dB, 6.32 dB and 5.73 dB for GC with compact taper, GC with adiabatic taper and focusing GC respectively. The insertion loss of the adiabatic long taper is slightly higher, which we can attribute to the waveguide loss in the adiabatic section.

Fig. 5(b) shows the response of the proposed taper and adiabatic taper by subtracting the patch response. Using the patch waveguides, after deducting the coupler loss, we observe insertion loss <0.7 dB and <0.8 dB per taper for a compact taper and adiabatic taper respectively. The performance of the proposed taper is marginally better than the adiabatic with 93% reduction in length.

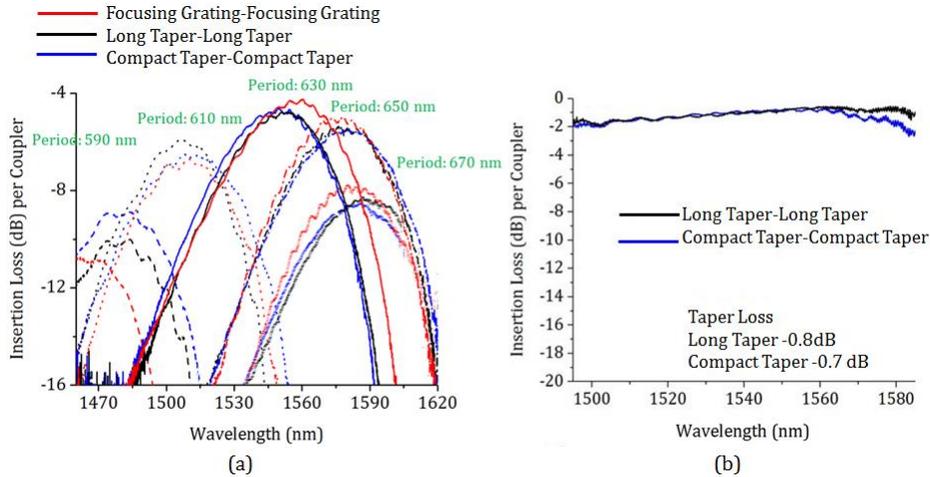

Fig. 5. Coupling efficiency of the various configuration of the GCs with variation in the period, (b) Insertion loss of the taper alone by neglecting the loss of the GC through a patch structure.

**Table 2. Characterization result of the proposed devices with different periods**

|  | Period (nm) | Coupling Efficiency per Coupler | 3 dB Bandwidth | Footprint |
|---|---|---|---|---|
| **Long Taper** | 590 | 9.9 | 50 | 210 |
|  | 610 | 5.9 | 52 |  |
|  | 630 | 6.32 | 54 |  |
|  | 650 | 6.9 | 52 |  |
|  | 670 | 8.3 | 56 |  |
| **Focusing Grating** | 590 | 10.9 | 50 | 40 |
|  | 610 | 6.6 | 50 |  |
|  | 630 | 5.73 | 53 |  |
|  | 650 | 6.49 | 50 |  |
|  | 670 | 7.8 | 52 |  |
| **Compact Taper** | 590 | 8.8 | 51 | 44.2 |
|  | 610 | 6.5 | 56 |  |
|  | 630 | 6.2 | 58 |  |
|  | 650 | 6.9 | 54 |  |
|  | 670 | 8.5 | 56 |  |

**Table 3. Change in efficiency with Rotational and Translational Misalignment for linear and focusing gratings**

|  | Rotational Misalignment | Insertion Loss (dB) | 3-dB Bandwidth | Translational Shift (pm) | Insertion Loss (dB) | 3-dB Bandwidth |
|---|---|---|---|---|---|---|
| **Linear Gratings** | $-10^0$ | 13.9 | 51 nm |  |  |  |
|  | -8 | 12.8 | 50 nm | -800 | 15.6 | 60 nm |
|  | -6 | 11.9 | 61 nm | -600 | 12.4 | 40 nm |
|  | -4 | 11.7 | 57 nm | -400 | 11.6 | 40 nm |
|  | -2 | 7.87 | 62 nm | -200 | 9.34 | 40 nm |
|  | **0** | **5.43** | **62 nm** | **0** | **9.0** | **41 nm** |
|  | 2 | 6.12 | 65 nm | 200 | 9.87 | 40 nm |
|  | 4 | 7.45 | 60.8 nm | 400 | 10.2 | 45 nm |
|  | 6 | 11.47 | 58.6 nm | 600 | 11.57 | 51 nm |
|  | 8 | 12.4 | 60.8 nm | 800 | 14.37 | 50 nm |
|  | 10 | 12.1 | 65 nm |  |  |  |
| **Focusing Gratings** | $-10^0$ | 19.1 | 53 nm |  |  |  |
|  | -8 | 21.0 | 65 nm | -800 | 19.07 | 70 nm |
|  | -6 | 18.1 | 68 nm | -600 | 14.4 | 58 nm |
|  | -4 | 13.2 | 54 nm | -400 | 10.97 | 44 nm |
|  | -2 | 7.61 | 58 nm | -200 | 8.7 | 40.6 nm |
|  | **0** | **5.29** | **60.1 nm** | **0** | **8.1** | **42 nm** |
|  | 2 | 6.1 | 58 nm | 200 | 8.89 | 41 nm |
|  | 4 | 10.56 | 56 nm | 400 | 10.9 | 44 nm |
|  | 6 | 16.9 | 42 nm | 600 | 14.26 | 51 nm |
|  | 8 | 19.2 | 48.7 nm | 800 | 18.03 | 61 nm |
|  | 10 | 18.3 | 52 nm |  |  |  |

Table 3 compares the alignment tolerance of the focusing and the linear GCs. To obtain the rotational alignment tolerance, the angle of the GCs were fabricated with a $\Delta$ shift of $2^0$ clockwise and anticlockwise (Fig. 6(a) and (b) shows a $2^0$ anticlockwise shift in grating placement). As is evident from Fig. 6(c), linear GCs are more tolerant with a roll-off of 2.613 dB/degree and 1.32 dB/degree (linear part) for focusing and linear gratings respectively.

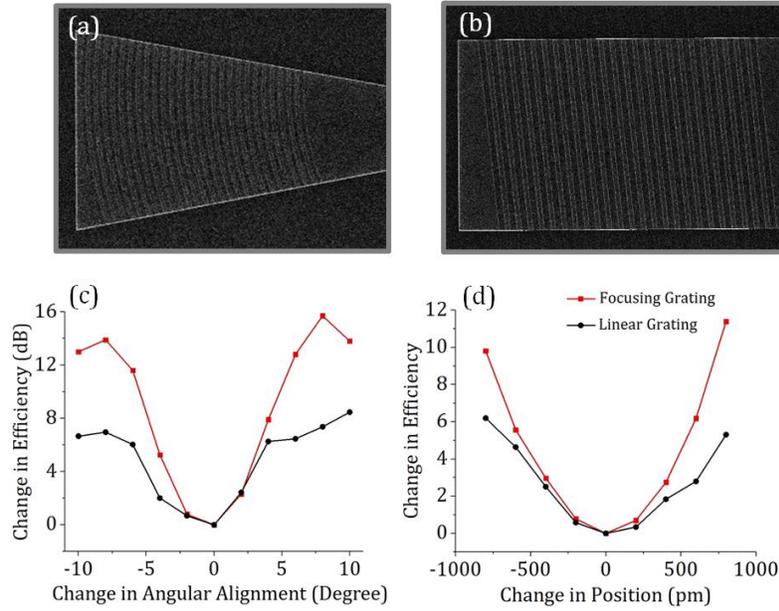

Fig. 6. (a) and (b) SEM image of the rotational misalignment in the focusing and linear gratings by $2^0$, Change in coupling efficiency with (c) Rotational misalignment and (d) Translational misalignment of the focusing and linear gratings.

Table 4. Various adiabatic and non-adiabatic grating assisted tapers on SOI proposed in literature

| | Taper Designs | Length (μm) | Initial Width (μm) -Final Width (μm) | Coupling Efficiency | Remarks |
|---|---|---|---|---|---|
| **Adiabatic** | Linear | 20-200 | 10 – 0.5 | 44.9-98.5% | Trade-off between the taper length and coupling efficiency due to the adiabatic transition |
| | Exponential (Positive) | 20-200 | 10 – 0.5 | 14-99.5% | |
| | Exponential (Negative) | 200 | 10 – 0.5 | 48.5-97.4% | |
| | Parabolic | 200 | 10 – 0.5 | 32.4-98.6% | |
| | Efficient Adiabatic[19] | 120 | 12 – 0.5 | 98.3% | |
| | Adiabatic Taper Based on Thin Flat Focusing Lenses[20] | 22.5 | 10 – 0.5 | 95.4% | Silicon subwavelength slits are fabrication alignment intolerant |
| | Hollow tapered spot-size converter[21] | 60 | 15 – 0.3 | 72% | |
| **Non-Adiabatic** | Segmented[22] | 15.4 | 10 – 0.56 | 98.3% | Complexity in fabrication |
| | Segmented- Stepwise New[23] | 20 | 12 – 0.5 | 92.1% | |
| | Lens-assisted[24] | 20 | 10 – 0.45 | 1 dB (TE), 5 dB (TM) (Experimental) | |
| | Discontinuous[25] | | 10 – 0.45 | 90% | |
| | **Proposed Taper** | **34.2** | **10 – 0.5** | **95%** | **Robust to Fabrication Errors, Broadband, Efficient** |

Vermeulen *et. al.* have shown tilted GCs to minimize Fresnel back-reflections by designing the grating teeth such that reflections are directed away from the aperture of the focusing GC [31]. However, we have arbitrarily varied the angle as in the case of overlay misalignment to inspect the robustness of the gratings. Fig. 6(d) compares the translational alignment tolerance of the focusing and linear GC where one of the fibers is aligned from the optimum position and the change in efficiency is measured. As is evident, focusing gratings are less tolerant to fiber alignment as compared to linear gratings. The roll-off for the linear part (since the graph is parabolic in nature) is 13.5 dB/fm and 8.3 dB/fm for focusing and linear GC respectively. The devices based on proposed compact tapers are 11 times smaller (~ 442 µm$^2$) in comparison to linear GC-based couplers (5100 µm$^2$). Table 4 summarizes the various configurations of adiabatic/non-adiabatic grating assisted tapers proposed in literature. The proposed tapers have been optimized for the TE polarization only. However, similar structures would work for TM coupling as well.

The compact lateral waveguide tapers are necessary to realize coupling between devices of varying dimensions. We have designed and demonstrated a compact tapered spot-size converter to couple light to a single mode waveguide from a 10 µm wide waveguide. By using the taper with a linear GC, we have experimentally shown no degradation in coupling efficiency compared to standard focusing GC. We have also shown that the proposed structure achieves an improved 3 dB bandwidth of ~58 nm against ~53 nm (focusing GC) in the 1550 nm band. The device shows 11X reduction in the footprint of a single device based on linear GCs using adiabatic tapers. We have also shown the fabrication tolerance of the compact taper by varying various parameters. Moreover, translation as well as rotational alignment tolerance of the focusing and linear GC are also compared.